\documentclass[reprint,aps,nofootinbib,preprintnumbers,amsmath,amssymb,11pt]{revtex4}
\usepackage{graphicx}
\usepackage{dcolumn}
\usepackage{bm}
\usepackage{natbib}
\usepackage{color}

\newcommand{\nc}{\newcommand}
\newcommand\fft[2]{\frac{#1}{#2}}
\newcommand\ft[2]{{\textstyle\frac{#1}{#2}}}
\newcommand\nn{{\nonumber}}
\newcommand{\beq}{\begin{equation}}
\newcommand{\eq}{\end{equation}}
\newcommand{\es}{\eta/s}
\newcommand{\etas}{\frac{\eta}{s}}
\newcommand{\half}{\frac{1}{2}}

\nc{\bea}{\begin{eqnarray}} \nc{\ea}{\end{eqnarray}} \nc{\be}{\begin{equation}} \nc{\ee}{\end{equation}} \nc{\barr}{\begin{array}}
\nc{\earr}{\end{array}}

\begin{document}

\title{Generating Temperature Flow for $\es$ with Higher Derivatives:\\ From Lifshitz to AdS}
\preprint{DAMTP-2011-104 \; MIFPA-11-53 \; NSF-KITP-11-251}

\author{Sera Cremonini$ ^{\,\clubsuit,\spadesuit}{}^*$ and Phillip Szepietowski$ ^{\diamondsuit}{}^\dagger$ \\[0.4cm]
\it $ ^\clubsuit$ Centre for Theoretical Cosmology, DAMTP, CMS,\\
\it University of Cambridge, Wilberforce Road, Cambridge, CB3 0WA, UK \\ [.5em]
\it $ ^\spadesuit$ George and Cynthia Mitchell Institute for Fundamental Physics and Astronomy\\
\it Texas A\&M University, College Station, TX 77843--4242, USA \\ [.5em]
\it $ ^\diamondsuit$\textit{Department of Physics, University of Virginia,\\
Box 400714, Charlottesville, VA 22904, USA} }

{\let\thefootnote\relax\footnotetext{$^{*}$sera@physics.tamu.edu \\ $^{\dagger}$pgs8b@virginia.edu}}

\date{\today}

\begin{abstract}
We consider charged dilatonic black branes in $AdS_5$ and examine the effects of perturbative
higher derivative corrections on the ratio of shear viscosity to entropy density $\es$ of the dual plasma.
The structure of $\es$ is controlled by the relative
hierarchy between the two scales in the plasma, the temperature and the chemical potential.
In this model the background near-horizon geometry interpolates between a Lifshitz-like brane at low
temperature, and an AdS brane at high temperatures -- with AdS asymptotics in both cases.
As a result, in this construction the viscosity to entropy ratio flows as a function of temperature,
from a value in the IR which is sensitive to the dynamical exponent $z$,
to the simple result expected for an AdS brane in the UV.
Coupling the scalar directly to the higher derivative terms generates additional temperature dependence,
and leads to a particularly interesting structure for $\es$ in the IR.
\end{abstract}

\maketitle
\newpage
\tableofcontents
\newpage

\section{Introduction}

Over the past few years, the holographic gauge-gravity duality has gone through a series of transformations, moving further away from its most
comfortable arena, that of theories with a large amount of supersymmetry. Top-down studies, based on string/M-theory constructions, have been met by a
number of bottom-up approaches, with applications ranging from QCD-like theories to condensed matter systems.
For example, holographic techniques have been applied to probing the transport properties of the strongly coupled
QCD quark gluon plasma (QGP), with the notable order of magnitude agreement of the shear viscosity to entropy density
ratio $\etas$ between RHIC (and now LHC) data and the holographic result first found in \cite{Kovtun:2004de}\footnote{For
applications to $\es$ see for example the review articles \cite{Son:2007vk,Schafer:2009dj,Cremonini:2011iq},
and for an exhaustive review of the applications to the QCD plasma see \cite{CasalderreySolana:2011us} and references therein.}.

In the context of applications to condensed matter physics, recently much focus has been on
non-relativistic systems undergoing quantum critical behavior, which often exhibit an anisotropic scaling symmetry of the type
\beq
\label{Lscaling}
t \rightarrow \lambda^z t \, , \quad \quad \vec{x} \rightarrow \lambda \vec{x} \, ,
\eq
with the dynamical exponent $z$ generically not equal to one.
On the gravity side of the duality, this scaling can be realized as an isometry of the metric
\beq
\label{LifMetric}
ds^2= r^{2z} dt^2 + r^2 d\vec{x}^{\,2} + \frac{dr^2}{r^2}
\eq
as long as one also sends $r \rightarrow \lambda^{-1} r$. %
For the case of pure AdS (in the Poincar\'{e} patch), the dynamical critical exponent is simply $z=1$, describing relativistic scaling.
Studies of holographic duals exhibiting non-relativistic symmetries were initiated in \cite{Son:2008ye,Balasubramanian:2008dm},
with the first examples of Lifshitz scaling found in \cite{Kachru:2008yh}. There exist several string theory constructions of
non-relativistic holographic duals, including many which exhibit Lifshitz scaling \cite{Herzog:2008wg,Maldacena:2008wh,Adams:2008wt,arXiv:0905.0688,arXiv:1005.3291,Donos:2010tu,arXiv:1009.3445,arXiv:1009.3805,arXiv:1102.1727,Cassani:2011sv,Halmagyi:2011xh,arXiv:1103.1279,Chemissany:2011mb,Amado:2011nd,arXiv:1105.3472,arXiv:1106.1637}.
There is by now an extensive literature on the subject, and we refer the reader to the review articles
\cite{Sachdev:2008ba,Hartnoll:2009sz,Herzog:2009xv,McGreevy:2009xe} for a more detailed treatment
of holographic applications to condensed matter physics.

Our interest in systems exhibiting the anisotropic scaling symmetry (\ref{Lscaling}) stems from a rather different reason.
We would like to understand what are the ingredients needed to generate temperature dependence for the shear viscosity
to entropy ratio $\es$ of strongly coupled field theories, within the framework of the holographic gauge/gravity duality.
It is by now well-known that in all gauge theories
with Einstein gravity duals\footnote{An exception is the case of anisotropic fluids.
A way to obtain a deviation from universality by breaking the rotational
symmetry spontaneously was shown in \cite{Natsuume:2010ky,Erdmenger:2010xm}.
Moreover, the resulting non-universal shear viscosity to entropy ratio
has been shown to exhibit an interesting temperature flow, without the need for higher derivative corrections.
This can be seen in \cite{Erdmenger:2010xm,arXiv:1109.4592,arXiv:1110.0007}
in the context of p-wave superfluids,
as well as in \cite{arXiv:1110.6825}, in the anisotropic axion-dilaton-gravity background of \cite{arXiv:1105.3472,arXiv:1106.1637}.
} one finds
the universal result $\es = 1/4\pi$ \cite{Policastro:2001yc,Buchel:2003tz}.
Moreover, as argued originally via the membrane paradigm in \cite{Iqbal:2008by},
$\es$ is given strictly by horizon quantities, a result which holds even in theories with higher derivatives
\cite{Myers:2009ij,Paulos:2009yk,Eling:2011ct}.
Although $\es$ does not run in any Wilsonian sense\footnote{
For recent attempts at refining notions of holographic RG flow, see \cite{Bredberg:2010ky,Nickel:2010pr,Heemskerk:2010hk,Faulkner:2010jy}.}
(the associated radial flow on the gravitational side is trivial, explaining why it can be extracted from horizon data)
it can still undergo \emph{temperature flow}.
As a concrete example of a setting where such a flow is generated,
we will consider a system which exhibits non-relativistic scaling at low temperatures, and standard
relativistic scaling in the high temperature regime.
We should note that having a handle on the behavior of $\es$ as a function of temperature
is relevant for the physics of the strongly coupled QGP, and in particular for better understanding
the elliptic flow measurements from heavy ion collisions at RHIC and LHC (see e.g. \cite{Shen:2011kn,Niemi:2011ix}).

Our starting point will be Einstein-Maxwell-dilaton gravity in five dimensions.
Thanks to the presence of the dilatonic scalar field,
finite temperature solutions in this theory (with a finite charge density)
exhibit Lifshitz-like scaling near the horizon, while asymptoting to AdS
at large distances \cite{Taylor:2008tg,Goldstein:2009cv,Chen:2010kn,Perlmutter:2010qu,
Bertoldi:2010ca,Bertoldi:2011zr,Gouteraux:2011ce,Berglund:2011cp,Gouteraux:2011qh}.
Of particular interest to us, however, is the temperature dependence of the solutions.
In fact, the Lifshitz-like scaling regime appears in the near-horizon geometry
when the temperature is much smaller than the chemical potential set by the finite charge density, $T\ll \mu$ (\emph{i.e.} in the IR).
On the other hand when $T \gg \mu$ (\emph{i.e.} in the UV), the near-horizon solution is replaced by an anti-de Sitter black brane.
The interpolation between the low-$T$ and high-$T$ regimes has been
shown to be smooth -- no discontinuous phase transition occurs as $T$ is varied \cite{Bertoldi:2010ca,Bertoldi:2011zr}.
An interesting temperature flow for the viscosity to entropy ratio is generated by adding appropriate higher derivative corrections to the theory
-- without which there would be no deviation from the universal $\es = 1/4\pi$ result.

Once such higher derivative terms are turned on, we will find that the behavior of $\es$ in the $T\ll\mu$ regime
is sensitive to the dynamical exponent $z$, while for $T\gg\mu$ it is entirely independent of it --
knowledge of $z$ has been washed out by the large temperature in the system.
Moreover, $\es$ will interpolate smoothly between its IR and UV values as the system is heated up.
While this is not Wilsonian RG flow, it does offer an example of a
setting in which the structure of $\es$ in the IR is very different from that in the UV,
and in which UV data alone is not enough to specify the behavior of $\es.$ 
Another example of such a stark difference between UV and IR physics
-- and of an analogous temperature flow --
was seen in \cite{Buchel:2010wf}, where the system underwent a second order (superfluid) phase transition\footnote{In the
construction of \cite{Buchel:2010wf} the higher derivative terms were engineered to be non-vanishing only \emph{below} the
temperature $T_c$ associated with the phase transition. As a result, in \cite{Buchel:2010wf} one finds the universal result
$\es=1/4\pi$ above $T_c$, while a deviation from it below $T_c$, with an associated running with temperature.}.
Just like in the model of \cite{Buchel:2010wf}, here the non-trivial behavior for the viscosity to entropy ratio can be traced to the presence
of a matter field, and in particular to its coupling to the higher derivative terms. In fact, as we will see, it will be the dependence of $\es$ on the (horizon value of the) scalar which will lead to
\emph{additional}
temperature dependence, and a significantly more intricate behavior in the IR.
Although a dependence of $\es$ on $T/\mu$ was already seen
in theories with higher derivative corrections at finite chemical
potential \cite{Myers:2009ij,Cremonini:2009sy}, its structure was not as rich as the one observed in this setup,
where it is sensitive to the dilatonic couplings and the different scalings in the low-$T$ and high-$T$ regimes.

To summarize, our main goal in this paper was to gain insight into what features are needed
to construct a system in which transport coefficients such as $\es$ undergo temperature flow.
The model we have chosen here provides a concrete example -- in no way unique -- of such a setting,
where the evolution of a scalar field translates into a flow for $\es$.
Interestingly, in this model a particularly non-trivial temperature dependence arises when a shift symmetry of the two-derivative
action is explicitly broken by the higher-derivative terms -- linking a much richer
behavior of the transport coefficients to the presence of broken symmetries.
In fact, breaking of this shift symmetry will be crucial for achieving a particularly significant deviation from the
behavior of the standard Reissner-Nordstr$\ddot{o}$m anti de-Sitter (RN-AdS) black brane background studied in \cite{Myers:2009ij,Cremonini:2009sy}.
Finally, we should mention that, although understanding the intermediate temperature regime (when $T \sim \mu$) in our model would be
interesting, such an analysis is beyond the scope of this paper, and we have restricted our attention to the endpoints of the flow.

The organization of the rest of the paper is as follows.
In Section \ref{section:setup} we introduce the Einstein-Maxwell-dilaton theory we will focus on, and
discuss relevant properties of its black brane solutions.
We present the details of the calculation of $\es$ in Section \ref{section:etascalc}.
We also include an expression for $\es$ in terms of a generic near-horizon black brane expansion, which will facilitate the discussion
of temperature dependence.
Section \ref{section:etasspecifics} is dedicated to the flow of $\es$ as a function of temperature.
We discuss our main results in Section \ref{Conclusions}.

\section{The Setup}\label{section:setup}

We take as a starting point the following Einstein-Maxwell-dilaton action,
\begin{equation}
\label{action0}
S_0 = \frac{1}{16 \, \pi G_{5}} \int d^{5}x\sqrt{-g} (R - 2\Lambda - 2 \left(\nabla \phi\right)^2 - e^{2\alpha\phi} F^2) \, ,
\end{equation}
where $\phi$ is a real scalar and $F = d\mathcal A$ is an abelian field strength. Thanks to the presence of the $U(1)$
gauge symmetry in the bulk, the dual field theory has
a conserved particle number, with an associated chemical potential $\mu$.
The action is also invariant under the shift symmetry
\beq
\label{shiftsymm}
\phi \rightarrow \phi+ \delta\, , \qquad \mathcal A \rightarrow e^{-\alpha \delta} \mathcal A \, ,
\eq
reflecting the fact that the normalization of the gauge coupling can be absorbed into the
definition of the gauge field (and in particular, in the definition of the charge).

Since we are interested in higher-derivative modifications to the shear viscosity to entropy density ratio, we will
add to the leading order action $S_0$ curvature corrections of the form
\begin{equation}
\label{action1}
S_1 = \frac{L^2}{16\pi G_5}\int d^{5}x\sqrt{-g}\left(\beta_1 \, e^{\,\gamma_1\phi}R_{\mu\nu\rho\sigma}R^{\mu\nu\rho\sigma}
+ \beta_2 \, e^{\, \gamma_2\phi}R_{\mu\nu\rho\sigma}F^{\mu\nu}F^{\rho\sigma} \right),
\end{equation}
with $\gamma_1,\gamma_2$ arbitrary constants.
We will take the couplings $\beta_1$ and $\beta_2$ of the higher derivative terms to be perturbatively small,
as would be the case in a derivative expansion within a genuine string theory reduction.
Note that generic scalar couplings $\gamma_i \neq 0$ explicitly break the shift symmetry (\ref{shiftsymm}) of the two-derivative action.
However, for the special case of $\gamma_1=0$ and $\gamma_2 = 2\alpha$ the symmetry is preserved.
Whether the shift symmetry is broken or not will play an interesting role in the discussion of $\es$ -- and in particular
of its temperature dependence -- as we will see in Section \ref{section:etasspecifics}.
Finally, although (\ref{action1}) is clearly not the most general form for the action at the four-derivative level,
the terms which we have included are the only ones that will contribute to $\es$, as we have checked explicitly\footnote{We note,
however, that we have not considered the addition of terms involving derivatives of the curvature, such as $e^{\gamma_3\phi}\nabla^2R$ or
$e^{\gamma_4\phi}\nabla^a\nabla^b R_{ab}$ (or terms related to these by integration by parts). Their inclusion is more subtle, since
the application of the Wald entropy formula in this case
becomes slightly more involved \cite{Jacobson:1993vj}.}
in accord with the general arguments put forth in \cite{Myers:2009ij}.

Before turning to the discussion of higher derivative corrections to $\es$, we would like to highlight some of the properties
of the solutions to the leading order two-derivative action $S_0$.
We will restrict our attention to electrically charged black brane solutions to (\ref{action0}), whose thermodynamics has been
investigated recently in a number of
studies \cite{Taylor:2008tg,Goldstein:2009cv,Chen:2010kn,Perlmutter:2010qu,Bertoldi:2010ca,
Bertoldi:2011zr,Gouteraux:2011ce,Berglund:2011cp,Gouteraux:2011qh}, aimed at applying holographic methods to strongly coupled condensed matter systems.\footnote{See also \cite{Dehghani:2010gn} for a discussion on the thermodynamics of Lifshitz geometries including higher derivatives.}
The finite temperature solution to (\ref{action0}) which is perhaps best known is the
uncharged AdS black brane,
\begin{eqnarray}
\label{adsbranesol}\label{AdSbrane}
ds^2 &=& -f(r) \, dt^2 +\frac{r^2}{L^2}\, d\vec{x}^2 + \frac{dr^2}{f(r)}\, ,
\quad\quad f(r)=\frac{r^2}{L^2}\left(1-\left(\frac{r_h}{r}\right)^4\right) \, ,\\
\phi(r) &=& \phi_0, \qquad  \mathcal A = A_0 \, dt \, ,
\end{eqnarray}
in which the gauge field and the dilatonic scalar field are both constant.
Here the radius $L$ is determined by the size of the cosmological constant, $\Lambda = -6/ L^2$. For this solution the temperature and entropy density are given by
\begin{equation}\label{AdSthermo}
T = \frac{r_h}{\pi L^2}, \qquad\qquad s = \frac{r_h^3}{4G_5L^3} = \frac{L^3\pi^3}{4G_5}\, T^3 \; .
\end{equation}

However, what we are interested in are solutions with a \emph{finite charge density}, and therefore a non-trivial gauge potential.
To this end, we note that the action (\ref{action0}) also admits a black brane solution with a metric exhibiting
the anisotropic Lifshitz scaling (\ref{Lscaling}), a non-trivial gauge field and a logarithmically running dilaton.
The scaling symmetry in this case is not exact, but is broken by the running of the dilaton --
as a result, the solution has been termed \emph{Lifshitz-like} \cite{Goldstein:2009cv}.
Taking the following ansatz for the fields,
\begin{eqnarray}
\label{generalansatz}
ds^2 = -a(r)^2 dt^2  + b(r)^2 d\vec{x}^2 + c(r)^2 dr^2  , && \nonumber \\
\phi = \phi(r) , \qquad \mathcal A = A(r) \, dt \, ,&&
\end{eqnarray}
the Lifshitz-like black brane solution to the action (\ref{action0}) is given by
\cite{Taylor:2008tg,Goldstein:2009cv,Chen:2010kn,Bertoldi:2011zr,Berglund:2011cp}:
\begin{eqnarray}
\label{Lifbranesol}
a(r)^2 = \Delta \, \frac{r^{2z}}{L^{2z}}\left(1-\left(\frac{r_h}{r}\right)^{z+3}\right), \qquad b(r)^2 = \frac{r^2}{L^2} , && \nonumber \\ c(r)^2 =
\Delta\frac{L^2}{ r^2}\left(1-\left(\frac{r_h}{r}\right)^{z+3}\right)^{-1}\!\!\!\!\!\!, \qquad e^{2\alpha\phi(r)} = \frac{L^6}{r^{6}} \Phi,  && \nonumber \\
A(r) = \frac{(z-1)L^2}{2Q}\frac{r^{z+3}}{L^{z+3}}\left(1 - \left(\frac{r_h}{r}\right)^{z+3}\right) \, . &&
\end{eqnarray}
The parameters specifying the solution are determined by the charge $Q$ of the $U(1)$ gauge field
and the dynamical exponent $z$, which is itself fixed by the value of the dilatonic coupling $\alpha:$
\begin{equation}
\Delta = \frac{(z+3)(z+2)}{12}, \qquad \Phi = \frac{Q^2(z+2)}{6 L^4(z-1)}, \qquad z = \frac{6 +\alpha^2}{\alpha^2} \, .
\end{equation}
Note that with our notation the dimension of the charge is $[Q]=L^2$.
Also, the temperature of this scaling solution is
\begin{equation}
T = \frac{(z+3)}{4\pi L^{z+1}} \, r_h^z\, ,
\end{equation}
with the appropriate units of energy, $[T]=1/L$.
Noting that the entropy density still behaves as $s \sim r_h^3$ and that the chemical potential also has units of energy $[\mu]=1/L$,
simple dimensional analysis tells us that the entropy must scale as
$s \sim T^{\frac{3}{z}} \mu^{3-\frac{3}{z}}$.

In the Lifshitz-like geometry the gauge field and the $e^{2\alpha\phi} F^2$ term in
the action diverge in the large $r$ region.
The divergence can be cured, however, by embedding the scaling solution (\ref{Lifbranesol})
into an asymptotically AdS geometry, with the embedding providing \emph{a relativistic UV completion} of the model.
The new geometry can be obtained by adding to the scaling solution (\ref{Lifbranesol}) a perturbation that is irrelevant close
to the horizon, but which becomes increasingly important for larger values of $r$.
In the new asymptotic region both the scalar and the gauge field approach constant values --
their divergence effectively cut-off by the scale at which the theory becomes AdS-like.
It is the chemical potential $\mu$ which plays the role of the cut-off, setting
the scale for the cross-over between the two distinct behaviors of the geometry.
Finite-temperature brane solutions interpolating between a near-horizon Lifshitz-like regime and an asymptotically AdS region -- realizing the
embedding described above -- have been constructed explicitly in \cite{Goldstein:2009cv,Chen:2010kn,Perlmutter:2010qu,Bertoldi:2010ca,Bertoldi:2011zr}.

It is important to emphasize that in the dilatonic system described by (\ref{action0}) the nature of the solutions is sensitive to the
hierarchy between the two scales in the theory, the temperature $T$ and the chemical potential $\mu$.
At a fixed temperature $T \ll \mu$, the near-horizon geometry $r \sim r_h$ is that of the Lifshitz-like brane (\ref{Lifbranesol}),
while the large $r$ behavior of the metric and gauge field is described by the \emph{same asymptotics} as those of
a planar charged AdS black hole, namely,
\bea
\label{chargedAdSasymptotics}
f(r) &=& \frac{r^2}{L^2}\left(1 - \frac{2M}{r^4} + \frac{Q^2e^{-2\alpha\phi_0}}{3r^6} + ... \right) \, ,\nn \\
A(r) &=&  \mu - \frac{Qe^{-2\alpha\phi_0}}{2Lr^2} + ...  \; , \nn \\
\phi(r) &=& \phi_0 + \frac{\phi_1}{r^4} + ... \; ,
\ea
where the ellipses denote terms of higher order in $1/r^2$.
The physical chemical potential $\mu$ can be read off as usual from the non-normalizable mode of the gauge field, and we have
allowed for an additional constant $e^{-\alpha \phi_0}$ for reasons which will become clear shortly\footnote{We could have absorbed it into
the definition of charge.}.
Away from the horizon and the boundary, the full geometry does not admit a simple analytical form and one must resort to
studying it numerically (we
refer the reader to \cite{Goldstein:2009cv,Perlmutter:2010qu,Bertoldi:2010ca,Bertoldi:2011zr} for a discussion of the numerics).

On the other hand,
when the temperature is the dominant scale in the system, $T \gg \mu$, the geometry is AdS-like everywhere.
In fact, when $T$ is so high that the charge is essentially negligible,
we can think of the uncharged AdS brane (\ref{adsbranesol}) as describing the entire geometry, from the horizon to the boundary.
With the exception of this very large temperature regime, the full solution for $T \gtrsim \mu$
should be thought of as a perturbation of (\ref{adsbranesol}),
whose behavior at intermediate distances is again only known numerically.
Solutions in the intermediate temperature region have been analyzed numerically in \cite{Bertoldi:2010ca,Bertoldi:2011zr},
where it was shown that the system evolves smoothly from the low-$T$ to the high-$T$ regime.

In summary, in the construction sketched above we see two types of `flows.'
One is the standard radial flow between a near-horizon geometry (Lifshitz-like if $T\ll \mu$)
and AdS asymptotics.
The other flow occurs as a function of temperature.
By heating up the system, the near-horizon Lifshitz brane which is present at $T\ll\mu$
is replaced by an AdS brane, with the chemical potential controlling the onset of the transition.
Once the chemical potential is negligible compared to the temperature of the system, $T\gg \mu$,
the AdS geometry is the one that dominates.
As we already mentioned above, the interpolation between the two opposite regimes is smooth \cite{Bertoldi:2010ca,Bertoldi:2011zr}.
It is the latter temperature flow that we will be mostly interested in, since it will provide us with precisely the type
of setup we need to probe the behavior of $\es$.
Also, while it would be interesting to study the system at arbitrary temperatures, here we will focus only on the two limiting cases
$T\ll\mu$ and $T\gg\mu$, which will be sufficient to illustrate our main points.

\section{Shear Viscosity to Entropy Ratio}\label{section:etascalc}
\label{Shear}

We are now ready to compute the viscosity to entropy ratio\footnote{Dilatonic couplings to higher derivative terms in
the context of $\es$ have also been explored in \cite{Cai:2009zv}. We should also point out that Gauss-Bonnet corrections to $\es$
in the model described by our two-derivative action (\ref{action0}) have been studied in \cite{Chen:2010kn} where, however, the higher derivative terms
were not coupled to the scalar.}
for the full theory
\bea
\label{fullaction}
S &=& \fft1{16\pi G_5} \int d^{5}x\sqrt{-g} \, \Bigl[R - 2\Lambda - 2 (\nabla\phi)^2 - e^{2\alpha\phi} F^2 \nn \\
&& \phantom{\fft1{16\pi G_5} \int d^{5}x\sqrt{-g} [}+ L^2
\left(\beta_1 \, e^{\,\gamma_1 \phi}\, R_{\mu\nu\rho\sigma}R^{\mu\nu\rho\sigma}
+ \beta_2 \, e^{\,\gamma_2 \phi}\, R_{\mu\nu\rho\sigma}F^{\mu\nu}F^{\rho\sigma} \right)
 \Bigr] \, ,
\ea
under the assumption that the higher derivative couplings $\beta_1$ and $\beta_2$ are perturbatively small.
Throughout the paper we will be working to linear order in the couplings $\beta_i$.
For reasons which will become clear shortly, for the $\es$ calculation it is particularly
convenient to make the coordinate change $u=r_h^2/r^2$ and rewrite
the black brane solution (\ref{Lifbranesol}) in the form
\bea
\label{convenientmetric}
ds^2 &=& - \Delta\frac{r_h^{2z}}{L^{2z}} \frac{f(u)}{u^z} \, dt^2 + \Delta \frac{L^2}{4u^2 f(u)} \, du^2 +
\frac{r_h^2}{L^2} \frac{1}{u} (dx^2 + dy^2 + dz^2) \, , \nonumber \\
\mathcal A &=& \frac{r_h^z}{L^z} \, h(u)\, dt, \qquad \quad \quad e^{2\alpha \phi(u)} = \frac{L^6 \, \Phi}{r_h^6} \, u^3 \, ,
\ea
where for the Lifshitz-like brane one has:
\beq
f(u) = 1 - u^{(z+3)/2} \,, \qquad \qquad
h(u) = \frac{(z-1) \, r_h^3}{2 L\, Q }\; u^{-(z+3)/2}f(u) \, .
\eq
The boundary and the horizon are now located at $u=0$ and $u=1$, respectively.
We note that the standard AdS brane solution (\ref{AdSbrane}) also fits into this form, with $z=1$ and the scalar
field and gauge field set to fixed values.

Although the metric (\ref{convenientmetric}) is a solution to the two-derivative theory (\ref{action0}) only,
we emphasize that for the computation of $\es$ knowledge of the back-reaction due to (\ref{action1}) is \emph{not} needed.
This is a simple consequence of the universality of the shear viscosity to entropy ratio
in asymptotically AdS backgrounds.
To see why this is so, recall that for Einstein gravity $\es=1/4\pi$, independently of the form of the solution.
The terms that would lead to a deviation from this universal result can only come from the higher-derivative
corrections in the action, which are already of order ${\cal O}(\beta_i$).
Thus, working to first order in the perturbative parameters $\beta_i$,
one only needs the background solution -- the back-reacted solution will yield a correction which is second order in perturbation theory.
In particular, this means that when computing $\es$
we don't have to take into account the renormalization of the dynamical exponent $z$ found in \cite{Adams:2008zk}.

In the framework of the gauge/gravity duality, the shear viscosity can be computed in a number of ways.
Here we will make use of Kubo's formula, which relates $\eta$
to the low frequency and zero momentum limits of the retarded Green's function of the CFT stress tensor:
\beq
\label{etaGreen}
\eta = - \lim_{\omega\rightarrow 0} \frac{1}{\omega} \, \text{Im} \,
G^R_{xy,xy}(\omega,\vec{k}=0) \, .
\eq
Let's sketch the main steps that go into finding $\eta/s$, following closely the discussion\footnote{Alternatively, $\eta/s$ can be extracted
from the effective graviton coupling \cite{arXiv:0808.3498,arXiv:0811.1665,arXiv:0901.3848}.} in \cite{Buchel:2004di,Myers:2009ij}.
The retarded Green's function can be extracted from the effective action $S_{eff}$ for the shear metric perturbation
$h_{\,x}^{\;\;\;y}(t,u) = \int \frac{d^4k}{(2\pi)^4} \, \phi_k (u) e^{-i\omega t + i k z}$.
Expanding (\ref{fullaction}) to quadratic order in the fluctuations $\phi_k(u)$ gives the
by-now standard form for the effective action \cite{Buchel:2004di},
\begin{eqnarray}
\label{effaction} S_{eff} & \sim \int \frac{d^4k}{(2\pi)^4} \; du & \Bigl[ A(u) \, \phi^{\prime\prime}_k \phi_{-k} +
B(u) \,\phi^{\prime}_k \phi^{\prime}_{-k} +
C(u) \, \phi^{\prime}_k \phi_{-k} + \nn \\ && + \; D(u) \, \phi_k \phi_{-k} + E(u) \, \phi^{\prime\prime}_k \phi^{\prime\prime}_{-k} + F(u) \,\phi^{\prime\prime}_k
\phi^{\prime}_{-k} \Bigr] + S_{GH} \, ,
\end{eqnarray}
where $S_{GH}$ denotes the (generalized) Gibbons-Hawking boundary term\footnote{For the explicit expression for
the Gibbons-Hawking boundary terms we refer the reader to
\cite{Buchel:2004di}.
For some of the subtleties involved in constructing
boundary terms and counterterms in theories with curvature-squared
corrections in the presence of a chemical potential see \emph{e.g.} \cite{Cremonini:2009ih}.}
and the coefficients $A(u), B(u),\ldots, F(u)$ encode information
about the background geometry.
After a series of manipulations the effective action can be shown to reduce to the boundary term
\beq
S_{eff} = \int \frac{d^4 k}{(2\pi)^4} \, \left. \mathcal{F}_k\right|_{u=0}^{u=1} \, ,
\eq
with the \emph{flux} term $\mathcal{F}_k$ given by
\beq
2 \mathcal{F}_k = \Pi_k \phi_{-k} + (C-A^\prime) \phi_k \phi_{-k} + E \phi_k^{\prime \prime} \phi_{-k}^\prime + \ldots \,,
\eq
where we have omitted a contribution coming from the
generalized Gibbons-Hawking boundary term\footnote{These terms will not be needed here.},
and $\Pi_k(u)$ is the radial momentum conjugate to $\phi_k$:
\beq
\label{momentum}
\Pi_k(u) \equiv \frac{\delta S_{eff}}{\delta \phi^\prime_{-k}} = \left( \bigl(B -A - \half F^\prime \bigr)\phi^{\prime}_k(u)
- \bigl(E \phi^{\prime\prime}_k(u)\bigr)^\prime \right) \, .
\eq
The retarded Green's function can be shown to be related to the flux through:
\beq
\label{green1}
G^R_{xy,xy} = -\lim_{u \rightarrow0} \frac{2\mathcal{F}_k}{\phi_k(u) \, \phi_{-k}(u)} \, .
\eq
Moreover, the only term in the flux that is relevant for the computation (\ref{etaGreen}) of the viscosity can be shown to be the first,
$2\mathcal{F}_k = \Pi_k \, \phi_{-k} + \ldots\;$. This, combined with (\ref{green1}), implies that
the Kubo relation (\ref{etaGreen}) reduces to the simple form
\beq
\label{eq:etapi}
\eta = \lim_{u,\omega\rightarrow0} \frac{\Pi_{\omega,k=0}(u)}{i\omega\phi_{\omega,k=0}(u)}\, ,
\eq
relating the shear viscosity to the boundary behavior of $\Pi_k$ and $\phi_k$, in the low-frequency limit \cite{Iqbal:2008by}.

However, one can do even better.
In terms of the radial momentum, the equation of motion for the fluctuations $\phi_k$ can be recast in the following way,
\beq
\partial_u\Pi_k(u) = M(u) \, \phi_k(u)\, ,
\eq
where $M(u)$ plays the role of an `effective mass' for the fluctuations, and is given by:
\beq
M(u) \equiv \frac{1}{16\pi G_5} \left(D - \half (C-A^\prime)^\prime \right) \, .
\eq
General arguments \cite{Myers:2009ij}
show that the effective mass must vanish in the low-frequency limit\footnote{Here we briefly
outline the argument of \cite{Myers:2009ij}. Consider the case of a constant $\phi(u)$,
for which the momentum $\Pi$ vanishes automatically, as clear from (\ref{momentum}).
Expanding the mass term as a series in the frequency, one can then show that
to satisfy the equation of motion we must have at least $M(u)={\cal O}(\omega)$.
Time reversal invariance then imposes that $M(u)={\cal O}(\omega^2)$, showing that the effective mass vanishes in the hydrodynamic limit.},
\beq
M(u) = \mathcal O (\omega^2) \, .
\eq
We have verified this explicitly for our four-derivative action (\ref{fullaction}) by evaluating $M(u)$ using the generic ansatz (\ref{generalansatz}),
confirming that in the hydrodynamic regime the momentum $\Pi_k(u)$ is independent of the radial direction $u$.
Similar arguments \cite{Myers:2009ij} show that, to leading order in the low-frequency limit, $\omega \, \phi(u)$ is also
independent of the radial position\footnote{At two-derivative level this follows from the fact that $\Pi$ is
held fixed in the low frequency limit and so $\omega\phi'(u) \propto \omega\Pi(u)$ vanishes.
The fact that it still holds in a perturbative expansion in the higher-derivative couplings simply follows
from the two-derivative result \cite{Myers:2009ij}.}.
Thus, we are free to evaluate (\ref{eq:etapi}) at any value of $u$, and in particular at the horizon radius, which simplifies greatly
the computation of $\es$ in many instances.
What we have seen, then, is that the shear viscosity is intrinsically a near-horizon quantity \cite{Iqbal:2008by}, even in theories with higher derivatives.

Putting all the ingredients together,
the final expression for $\eta$ given a metric of the form (\ref{convenientmetric}) and the effective action (\ref{effaction})
can be written \cite{Myers:2009ij} in the compact form
\beq
\label{prescription}
\eta = \frac{1}{8\pi G_5} \Bigl(\kappa_2(u_h) +\kappa_4(u_h)\Bigr)\, ,
\eq
where $u_h$ denotes the horizon radius, and
\bea
\kappa_2(u) &=& \sqrt{-\frac{g_{uu}(u)}{g_{tt}(u)}} \left(A(u)-B(u)+\frac{F^\prime(u)}{2}\right) \, , \\
\kappa_4(u) &=& \left(E(u) \left( \sqrt{-\frac{g_{uu}}{g_{tt}}} \right)^\prime \right)^\prime \,.
\ea
For our black brane metric (\ref{convenientmetric}) we arrive at:
\begin{equation}\label{FinalShear}
\eta = \frac{r_h^3}{16\pi G_5 L^3}\Biggl\{1 + \frac{2}{\Delta L^2}e^{\gamma_1\phi_h}\beta_1\left[\bigg(2(3z-1) - 4\gamma_1\phi^\prime (u_h)\bigg)
f^\prime(u_h)-4f^{\prime\prime} (u_h)\right]\Biggr\} \, .
\end{equation}
To find the entropy we make use of the standard Wald entropy formula \cite{Wald:1993nt},
\begin{equation}
\label{eq:Wald}
S = - 2 \pi \int_{\Sigma} d^{3} x \sqrt{- h} \frac{\delta {\mathcal L}}{\delta R_{\mu \nu \rho \sigma}} \,
\epsilon_{\mu \nu} \epsilon_{\rho \sigma} \, ,
\end{equation}
with $h$ the induced metric on the horizon cross section $\Sigma$, and $\epsilon_{\mu \nu}$ the binormal to $\Sigma$.
For our action and for the metric (\ref{convenientmetric}) we find that the entropy density $s = S/\omega_3$ is
\bea
\label{FinalEntropy}
s &=&
\frac{r_h^3}{4G_5L^3}\bigg\{1 +  \frac{2}{\Delta L^2}\Big[\big(2(3z-2)f'(u_h)-
  4f''(u_h)\big)e^{\,\gamma_1\phi_h}\beta_1 + \frac{8}{\Delta} \, (h'(u_h))^2 \,  e^{\,\gamma_2\phi_h}\beta_2 \Big]\bigg\},
\ea
where $\omega_3$ denotes the (infinite) black brane area.
Finally, the expression for the ratio of shear viscosity to entropy density is
\bea
\frac{\eta}{s} &=& \frac{1}{4\pi}\bigg[1 +  \frac{2}{\Delta L^2}\Big\{\big(2 - 4\gamma_1\phi'(u_h)\big)f'(u_h)
e^{\,\gamma_1\phi_h}\beta_1 - \frac{8}{\Delta} \, (h'(u_h))^2 \,  e^{\,\gamma_2\phi_h}\beta_2 \Big\}\bigg] \, ,
\ea
parametrized in terms of the values of $f(u), h(u)$ and $\phi(u)$ at the horizon,
as well as the couplings $\left\{\beta_i,\gamma_i \right\}$ of the higher-derivative corrections in the theory.

\subsection{Computing $\eta/s$ for a generic near-horizon solution}
\label{nearhorizonexp}

Given that the shear viscosity is a horizon quantity, as we have discussed above, we would like to close this section by
showing how $\es$ depends on
the parameters which specify the horizon data of a black brane solution\footnote{For a related discussion see \cite{Banerjee:2009ju}.}.
This will prove useful in the next section, once we discuss the behavior of $\es$ as one varies the temperature of the system.
With this motivation in mind, we start by writing down a generic near-horizon expansion for our scaling solution,
assuming a first order zero in $g_{tt}$ and a corresponding first order pole in $g_{rr}$,
and imposing regularity for the gauge field at the horizon.
Under these assumptions, the near horizon expansion is then of the form
\begin{eqnarray}
a(u)^2 &=& a_0(1-u) + a_1(1-u)^2 + a_2(1-u)^3 + ...\,, \nn \\
b(u)^2 &=& b_0  \,, \nn \\
c(u)^2 &=& c_0(1-u)^{-1} + c_1 + c_2(1-u) + ...\,, \nn \\
A(u) &=& A_0(1-u) + A_1(1-u)^2 + ...\,, \nn \\
\phi(u) &=& \phi_h + \phi_1(1-u) + \phi_2(1-u)^2 + ...\; \; .
\end{eqnarray}
Feeding this type of expansion into the prescription (\ref{prescription}) we find that the shear viscosity is
\begin{equation}
\eta = \frac{b_0^{3/2}}{16\pi G_5} \left(1 -
\frac{(a_0c_0 - a_0c_1 + 3c_0a_1 + 2a_0c_0\phi_1\gamma_1)}{a_0c_0^2}\, L^2\beta_1e^{\,\gamma_1\phi_h}\right)\, ,
\end{equation}
unaffected by the $R_{\mu\nu\rho\sigma} F^{\mu\nu} F^{\rho\sigma}$ term in the action, in agreement with \cite{Myers:2009ij,Cremonini:2009sy}.
The entropy density can be shown to be given by:
\begin{equation}
s = \frac{b_0^{3/2}}{4 G_5}\left(1 - \frac{(3c_0a_1 - a_0c_1)}{a_0c_0^2}\, L^2\beta_1e^{\,\gamma_1\phi_h} -
\frac{2A_0^2}{a_0c_0}\, L^2\beta_2e^{\,\gamma_2\phi_h}\right)\, .
\end{equation}
Combining these two expressions, we find that the viscosity to entropy density ratio reduces to:
\begin{equation}
\label{etasNH}
\etas = \frac{1}{4\pi}\left(1 - \frac{(1 + 2\gamma_1 \phi_1)}{c_0}L^2\beta_1e^{\,\gamma_1\phi_h} +
\frac{2A_0^2}{a_0c_0}L^2\beta_2e^{\,\gamma_2\phi_h}\right) \, .
\end{equation}

We can now evaluate (\ref{etasNH}) for the Lifshitz-like solution, for which the relevant near-horizon
data is given by
\begin{eqnarray}
\label{LifParams}
a_0 &=&\frac{1}{24}(z+2)(z+3)^2\frac{r_h^{2z}}{L^{2z}} \, , \quad\quad\quad c_0 = \frac{L^2}{24}(z+2) \, , \nn \\
A_0 &=& \frac{r_h^{z+3}}{4QL^{z+1}}(z-1)(z+3)\, , \qquad \quad
\phi_h = \frac{1}{2\alpha} \ln{\left(\frac{L^6\Phi}{r_h^6}\right)}\, , \qquad \phi_1 = \frac{3}{2\alpha},
\end{eqnarray}
and we recall that $\Phi = \frac{Q^2(z+2)}{6L^4(z-1)}$ and $z = \frac{6 +\alpha^2}{\alpha^2}$.
We find that the viscosity to entropy ratio for the scaling solution\footnote{In the absence of higher derivative corrections, the
result $\etas =\frac{1}{4\pi}$ had already been obtained
for systems with Lifshitz scaling at extremality in \cite{Imeroni:2009cs}.} becomes:
\begin{equation}\label{etasLifGeneral}
\frac{\eta}{s} = \frac{1}{4\pi}\left[1 - \frac{24(1-\frac{3\gamma_1}{\alpha})}{z+2}\; \beta_1 e^{\,\gamma_1\phi_h} +
 \frac{12(z-1)}{\, z+2} \; \beta_2 e^{(\gamma_2-2\alpha) \phi_h} \right].
\end{equation}

Let's analyze this result for the simplest case where we fix the dilaton couplings such that the global symmetry (\ref{shiftsymm})
is preserved, \emph{i.e.} $\gamma_1 = 0$ and $\gamma_2 = 2\alpha.$ In this case (\ref{etasLifGeneral}) reduces to
\begin{equation}
\label{etasLifShiftSymmetric}
\frac{\eta}{s} = \frac{1}{4\pi}\left(1 - \frac{24}{z+2}\, \beta_1 + \frac{12\, (z-1)}{z+2} \, \beta_2 \right).
\end{equation}
Thus, for this special choice of $\gamma_1$ and $\gamma_2$ the correction to $\es$ takes on a particularly simple form\footnote{As a
consistency check, we note that the $\beta_1$ term here agrees with the result of \cite{Chen:2010kn}, which included only Gauss-Bonnet corrections
and did not couple them to the scalar.}
dictated
by the value of the dynamical exponent $z$, which we recall is determined by the dilatonic coupling $\alpha.$
Note that when $z=1$, (\ref{etasLifShiftSymmetric}) reduces to the well-known case \cite{Kats:2007mq,Brigante:2007nu}
of an uncharged AdS brane,
\beq
\label{KP}
\etas = \frac{1}{4\pi} (1-8\beta_1) \, .
\eq
In that setup, the coupling $\beta_1$ of the higher derivative terms could be related to the
central charges $a$ and $c$ of the dual 4D CFT by making use of the trace anomaly\footnote{Similar arguments using the R-current anomaly relate $\beta_2$ to $a$ and $c$ in a very similar way.}, which gave $\beta_1 \propto \frac{c-a}{a}$.
Thus, (\ref{KP}) could be expressed \emph{entirely} in terms of the parameters of the UV CFT.
For the present case, in (\ref{etasLifShiftSymmetric}) we see that $\es$ is no longer uniquely determined by the UV central
charges (through its dependence on $\beta_i$) , but it is also sensitive to $z$, which parametrizes IR data.
Thanks to the different scalings of the solutions in our model, the expression for $\es$ in (\ref{etasLifShiftSymmetric}) is
explicitly sensitive to both the IR and the UV data. We will come back to this point in the next section.

Another interesting feature becomes apparent when one notices that the dependence on $z$ can be removed by making use of the
expression (\ref{LifParams}) for the scalar field at the
horizon\footnote{Using $e^{2\alpha\phi_h}=\frac{L^6 \Phi}{r_h^6}$ and recalling the form of $\Phi$, one finds
$$z = \left(1 + \frac{Q^2L^2e^{-2\alpha\phi_h}}{3r_h^6}\right)/ \left(1 - \frac{Q^2L^2e^{-2\alpha\phi_h}}{6r_h^6}\right) \, .$$
Note that in order to have $z>0$ we need $Q^2 < 6 r_h^6 e^{2\alpha\phi_h}/L^2$.}.
Trading the explicit $z$ dependence
for $\{Q, r_h, e^{\,\alpha \, \phi_h} \}$,
we find that (\ref{etasLifShiftSymmetric}) becomes:
\begin{equation}\label{etaschbb}
\frac{\eta}{s} = \frac{1}{4\pi}\left[1 - 4\left(2\beta_1 - (\beta_1 + \ft{3}{2} \beta_2)\frac{Q^2L^2e^{-2\alpha\phi_h}}{3r_h^6} \right)\right].
\end{equation}
With the charge rescaled appropriately, $\tilde Q = \frac{QLe^{-\alpha\phi_h}}{\sqrt{3}}$, the expression (\ref{etaschbb}) becomes
\emph{identical} in structure\footnote{
We stress however, that even though the form of $\es$ in (\ref{etaschbb}) is
suggestive, the charge $\tilde{Q}$ of the RN-AdS brane
would not be the same as the charge $Q$ measured at infinity from the solution with the near-horizon Lifshitz scaling.}
to the standard result\footnote{Note that our gauge field normalization differs by a factor of $1/4$ from that of
\cite{Myers:2009ij,Cremonini:2009sy}.}
 for a RN-AdS brane (with charge $\tilde Q$) in a theory with higher derivative corrections of the form
$\delta\mathcal{L} = \beta_1 R_{\mu\nu\rho\sigma} R^{\mu\nu\rho\sigma} +
\beta_2 R_{\mu\nu\rho\sigma} F^{\mu\nu}F^{\rho\sigma} + \ldots$, originally found in \cite{Myers:2009ij,Cremonini:2009sy}.
This is not too surprising, however, if we recall that $\es$ is a horizon quantity, and at the horizon the role of the dilatonic gauge kinetic coupling
$e^{2\alpha\phi}$ is essentially to rescale the charge.
We emphasize that this is a consequence of having chosen the dilatonic couplings in such a way to preserve the shift symmetry.
Thus, even though the low-$T$ Lifshitz solution of our dilatonic system is entirely different from the RN-AdS brane
of Einstein-Maxwell theory, the structure of $\es$ -- as long as the shift symmetry is not broken -- will be the same.

Finally, we would like to comment briefly on ramifications for the KSS bound.
Assuming the $\beta_i$ coefficients are strictly positive, we see that for the case where the dilaton coupling respects the shift symmetry,
as in (\ref{etasLifShiftSymmetric}), the $\beta_1$ term always decreases the value $\es$ whereas the $\beta_2$ term always increases it.
Moreover, as we take $z$ to be very large the $\beta_1$ contribution becomes vanishingly small while the
$\beta_2$ term approaches a constant value.
Thus, in the large $z$ limit, the $R_{\mu\nu\rho\sigma}F^{\mu\nu}F^{\rho\sigma}$ term dominates the correction.
The same behavior
was also observed in the RN-AdS black brane correction when evaluated at extremality
\cite{Myers:2009ij,Cremonini:2009sy}.

We close this section by commenting briefly on the case where the shift symmetry is explicitly broken, and
both $\gamma_1$ and $\gamma_2$ are allowed to take arbitrary values.
By examining (\ref{etasLifGeneral}) we see that turning on $\gamma_1$ gives additional freedom to change the sign of the correction
-- and the $\beta_1$ contribution can naively be forced to vanish by appropriately tuning $\gamma_1$.
Clearly the size and sign of the correction to the universal $1/4\pi$ result will now depend on the ranges (and signs) of all the parameters
$\{\gamma_i, \beta_i\}$ controlling the higher derivative terms, as well as on the value of the dynamical exponent.
Thus, once the shift symmetry is broken, the viscosity to entropy ratio extracted from the scaling solution
(\ref{etasLifGen}) is no longer related in any simple way to the value it would take for an AdS brane.
As we will see in the following section, this is tied to the fact that in this model $\es$ not only runs as a function of temperature
from its IR value to that in the UV, but can also have additional dependence on $T$ within the $T \ll \mu$ regime itself.

\section{Temperature Flow: from Lifshitz to AdS}\label{section:etasspecifics}

We are now ready to discuss the behavior of $\es$ as a function of temperature, and in particular in the two opposite regimes
$T\ll\mu$ and $T\gg\mu$, the IR and the UV respectively.
To get a better feel for the structure of $\es$, let's start by assuming that the scalar field does not couple
directly to the higher derivative corrections, $\gamma_i=0$, and turn off the $R_{\mu\nu\rho\sigma} F^{\mu\nu} F^{\rho\sigma}$  term
by setting $\beta_2=0$,
so that the global symmetry (\ref{shiftsymm}) of the action is restored.
We then see from (\ref{etasNH}) that $\es$ can be written in the simple form
\begin{equation}
\label{etasNH2}
\etas = \frac{1}{4\pi}\left[1 - \frac{L^2\beta_1}{c_0}\right] \, ,
\end{equation}
with the ${\cal O}(\beta_1)$ higher derivative correction parametrized entirely in terms of $c_0$ -- different near-horizon geometries
described by different values of $c_0$.
Moreover, since the near-horizon behavior of the solutions in our model
changes as $T$ and $\mu$ are varied,
(\ref{etasNH2}) shows in a clean way that the viscosity to entropy ratio will flow as a function of $T/\mu$
-- the temperature dependence encoded in $c_0$ translating into a flow for $\es$.
Let's describe this simple flow in a bit more detail.


At low temperatures $T\ll\mu$ the near-horizon geometry probed by the shear fluctuation
is that of the Lifshitz-like brane.
Reading off $c_0$ from (\ref{LifParams}), equation (\ref{etasNH2}) tells us that
in the IR the viscosity to entropy ratio has the remarkably simple form
\beq
\label{etasLifsimple}
\left . \etas \, \right\vert_{\text{IR}} = \frac{1}{4\pi}\left[1 - \frac{24}{z+2}\, \beta_1\right] \, , 
\eq
which we could have equivalently expressed\footnote{Trading $z$ for the coupling $\alpha$
we have $\etas = \frac{1}{4\pi}\left[1 - 8\, \frac{\alpha^2}{\alpha^2+2}\, \beta_1\right]$.}
in terms of the dilatonic coupling $\alpha$.
On the other hand, for temperatures much higher than the chemical potential, $T\gg\mu$,
the near-horizon geometry becomes the AdS-black brane --
the high temperature has essentially `washed-out' any footprints of the Lifshitz region in the IR.
In particular, when the temperature is so high that $\mu$ is negligible, the near-horizon behavior is that of an uncharged black
brane in AdS, for which both the gauge field and the scalar are constants. For this geometry $c_0 = L^2/8$, and (\ref{etasNH2}) reduces to
\beq
\label{uncharged}
\left . \etas \, \right\vert_{\text{UV}} = \frac{1}{4\pi}\left[1 - 8\beta_1\right] \, ,
\eq
in agreement with the well-known result of \cite{Kats:2007mq,Brigante:2007nu}.
As a simple consistency check, note that (\ref{etasLifsimple}) reduces to the uncharged AdS brane result (\ref{uncharged})
when the dynamical exponent becomes $z=1$, as it should.
Finally, when the temperature is large compared to $\mu$, but not so large that $\mu$ should be neglected,
the geometry starts being sensitive to the presence of charge.
Unfortunately, due to the non-vanishing scalar field in our system, the usual RN-AdS black brane is not a solution
of the theory, and in this intermediate temperature regime the near-horizon behavior is not known analytically.
As a result, $\es$ must be computed numerically.
Understanding in detail this middle region is, however, beyond the scope of this work.

Although we don't have an analytic expression for the near horizon geometry for arbitrary temperatures, we emphasize that
the parameter $c_0$ varies smoothly (and monotonically) \cite{Bertoldi:2011zr} between the value it takes in the IR and that in the UV,
\beq
\frac{L^2}{8} \leq c_0 \leq  \frac{L^2}{24}(z+2) \, ,
\eq
implying a smooth interpolation for $\es$ as a function of temperature\footnote{Note that this behavior is similar in spirit to
the temperature flow seen in higher derivative theories at finite
chemical potential \cite{Myers:2009ij,Cremonini:2009sy}.} :
\beq
\label{range}
\frac{1}{4\pi}\left[1 - 8\beta_1\right]  \leq \etas \leq \frac{1}{4\pi}\left[1 - \frac{24}{z+2}\, \beta_1\right] \, .
\eq
As we mentioned in Section \ref{section:etascalc},
the holographic Weyl anomaly relates
the coupling of the curvature-squared correction
$\sim \beta_1 R_{\mu\nu\rho\sigma} R^{\mu\nu\rho\sigma}$ to the central charges $a,c$ of the dual 4D CFT:
\beq
\label{centralcharges}
\beta_1 \propto \frac{c-a}{a} \, .
\eq
Thus, comparing the left- and right-hand sides of  (\ref{range}) we see that,
while at high temperatures $\es$ can be expressed \emph{entirely} in terms of the UV central charges $\{a,c\}$,
for low temperatures the correction is `dressed' by the dynamical exponent $z$, and is therefore also sensitive to IR data.
This feature sets this construction apart from the setups of \emph{e.g.} \cite{Kats:2007mq,Brigante:2007nu},
where UV data was enough to specify $\es$.
A similar behavior for transport -- with a sharply different $\es$ in the IR and in the UV --
was also found in \cite{Buchel:2010wf}, where it was intimately tied to the presence of a superfluid phase transition.

Given the relation (\ref{centralcharges}), it is natural to try to describe the flow (\ref{range}) at arbitrary temperatures
in terms of some combination $\mathcal C_{eff}$ of \emph{effective} central charges, encoding information about the number of
degrees of freedom of the dual system:
\beq
\etas = \frac{1}{4\pi}\left[1 - \mathcal C_{eff}\left(\frac{T}{\mu}\right) \right]\, .
\eq
Clearly, this function would have to reduce to our results at the endpoints of the flow,
\begin{equation}
\mathcal C_{eff}\left(\frac{T}{\mu}\right) \;\; \propto \
\begin{cases}
8\beta_1 \, , \qquad \qquad \; {\rm T\gg\mu } \; ,\\
\frac{24}{z+2} \, \beta_1 \, \,,\qquad \;  \; \; {\rm T\ll\mu }\; ,
\end{cases}
\label{fCeff}
\end{equation}
where we can associate $\beta_1$ with the appropriate central charges of the dual UV CFT.
Notice that, as long as $z>1$, the deep IR value (\ref{etasLifsimple}) of $\es$ is always larger\footnote{It is
largest in the $z\rightarrow \infty$ limit, in which it approaches $1/4\pi$.}
than the result (\ref{uncharged}) in the UV, with a monotonic flow between the two.
Thus, at least in this simple case, the size of $\mathcal C_{eff}$
decreases with temperature from the UV to the IR,
suggesting that it may in fact be a measure of the degrees of freedom in the system. It would be interesting to make this type of relation more precise.

Let's now go back to discussing the generic higher derivative action (\ref{action1}), with all the couplings $\{\beta_i, \gamma_i \}$ turned on.
First we note that for $T\gg\mu$, corresponding to the case where the near-horizon is AdS-like,
the correction to $\es$ is only slightly changed from that in (\ref{uncharged}):
\beq
\left . \etas \, \right\vert_{\text{UV}} = \frac{1}{4\pi}\left[1 - 8 \beta_1e^{\gamma_1\phi_0} \right] \, ,
\eq
where $\phi_0$ is the boundary value of the (constant) dilaton, as given in (\ref{chargedAdSasymptotics}).
We therefore focus the rest of the discussion on the more interesting case of the low temperature Lifshitz-like brane.
For convenience, we repeat here the full expression for $\es$ in the $T\ll\mu$ regime,
\begin{equation}
\label{etasLifGen}
\left . \frac{\eta}{s} \, \right\vert_{IR} = \frac{1}{4\pi}\left[1 - \frac{24(1-\frac{3\gamma_1}{\alpha})}{z+2}\; \beta_1 e^{\,\gamma_1\phi_h} +
 \frac{12(z-1)}{\, z+2} \; \beta_2 e^{(\gamma_2-2\alpha) \phi_h} \right],
\end{equation}
with arbitrary couplings, breaking the shift symmetry of the two-derivative action.
Compared to what we had in (\ref{etasLifsimple}), note that $\es$ is now sensitive to the horizon value $\phi_h$ of the scalar.
In the Lifshitz-like regime, the gauge kinetic coupling at the horizon
scales with temperature as
\beq
e^{-2\alpha\phi_h} \sim T^{\frac{6}{z}} \, .
\eq
Since $\mu$ is the only other scale in the theory,  simple dimensional analysis tells us that
\begin{equation}
\label{phi0scaling}
e^{-2\alpha\phi_h} \sim \left(\frac{T}{\mu}\right)^{6/z}
\end{equation}
is the appropriate scaling of the dilatonic factor with $T$ and $\mu$.
For the precise proportionality constant, we would need
the exact dependence of $\mu$ on the parameters of the near-horizon IR solution, which in our model is not known analytically
\cite{Bertoldi:2010ca,Bertoldi:2011zr}.

However, from the scaling (\ref{phi0scaling}) we can see that the dilatonic factors $e^{\gamma_i \phi_h}$ and $e^{-2\alpha \phi_h}$ in
(\ref{etasLifGen}) will lead to additional
temperature dependence in the IR, and give rise to a non-trivial behavior for $\es$.
Even in the absence of direct couplings of the scalar to the higher derivative terms (\emph{i.e.} $\gamma_i=0$), and setting $\beta_1=0$
for simplicity, we already see
a new temperature-dependent contribution:
\beq
\label{newT}
\left . \frac{\eta}{s}  \, \right\vert_{IR} - \frac{1}{4\pi} \; \; \propto \; \; \beta_2
\, \left(\frac{T}{\mu}\right)^{6/z} \, .
\eq
The simple $\beta_2=0$ limit of (\ref{etasLifGen}) also leads to additional temperature dependence, but is
a bit more subtle.
In this case the deviation of $\es$ from $1/4\pi$ becomes:
\beq
\label{diff}
\left . \frac{\eta}{s}  \, \right\vert_{IR} - \frac{1}{4\pi} \; \; \propto \; \; \beta_1\left(\frac{T}{\mu}\right)^{-\frac{\gamma_1}{z} \sqrt{3(z-1)/2}} \, .
\eq

For positive $\gamma_1$ this result should not be trusted at very small temperatures -- near extremality the correction
stops being perturbatively small, signaling a break-down of our analysis.
This behavior can be traced back to the dependence of the scalar field at the horizon on the temperature of the system.
In fact, divergences in the extremal limit of Lifshitz solutions are well-known (see
\cite{Copsey:2010ya,Horowitz:2011gh} for a recent discussion of stability issues of Lifshitz backgrounds), and higher derivative corrections
are expected to become large in the zero temperature limit.
A more quantitative analysis of the break-down of (\ref{diff}) would require a careful study of the solutions
in the regime of nearly zero temperature, as well as of the backreaction of the higher derivative terms.

Here we will content ourselves with pointing out that the presence of generic $\phi_h$ dependence in $\es$ leads to an additional sensitivity to temperature, even in the low-$T$ regime,
which one would \emph{not} see if the global symmetry was preserved.
In fact, contrast the cases we have just discussed to the situation with
$\gamma_1 = 0$ and $\gamma_2 = 2\alpha.$ Now the shift symmetry is intact,
any dependence of (\ref{etasLifGen}) on $\phi_h$ is lost and the IR expression for $\es$ reduces to
\begin{equation}\label{etasLifShiftSymm}
\left . \frac{\eta}{s} \right \vert_{IR}= \frac{1}{4\pi}\left(1 - \frac{24}{z+2}\, \beta_1 + \frac{12\, (z-1)}{ z+2} \, \beta_2 \right) \, ,
\end{equation}
with no trace left of the additional running with $T$.
In summary, much like in \cite{Buchel:2010wf}, it is the breaking of a symmetry (a shift symmetry here) which plays a crucial role
in generating a particularly non-trivial structure for the viscosity to entropy ratio.

\begin{figure*}[t]\label{fig:etas-simple}
\centering
\includegraphics[scale=0.6]{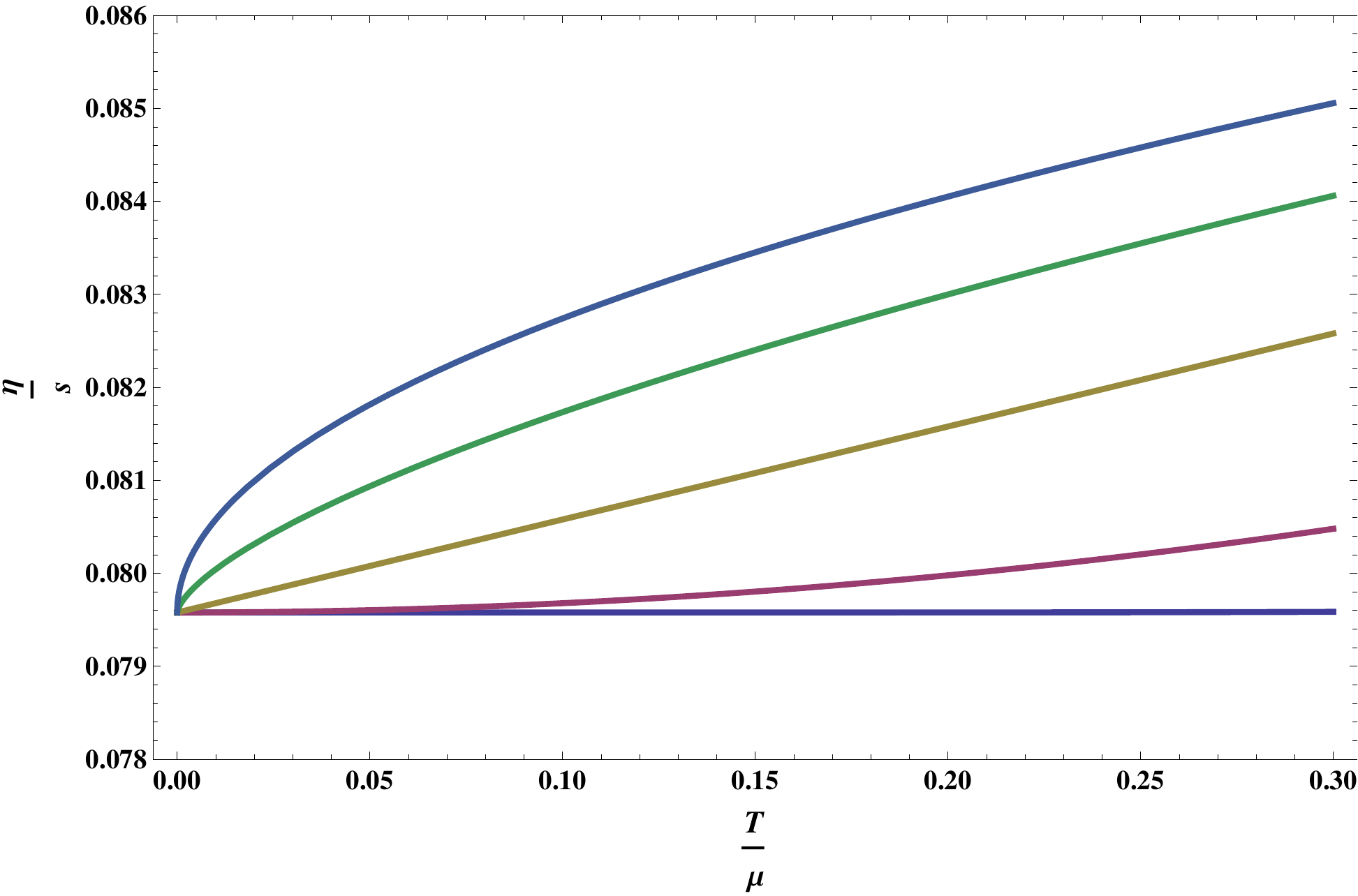}
\caption{$T/\mu$ dependence of $\eta/s$ for a simple case with $\{\beta_1 = \gamma_1 = \gamma_2 = 0\}$ and $\beta_2=0.01$.
The plots are for values of $z = \{12,\,9,\,6,\,3,\,1\},$ from top to bottom.}
\label{fig:etas-simple}
\end{figure*}

\begin{figure*}[t]
\centering
\includegraphics[scale=0.6]{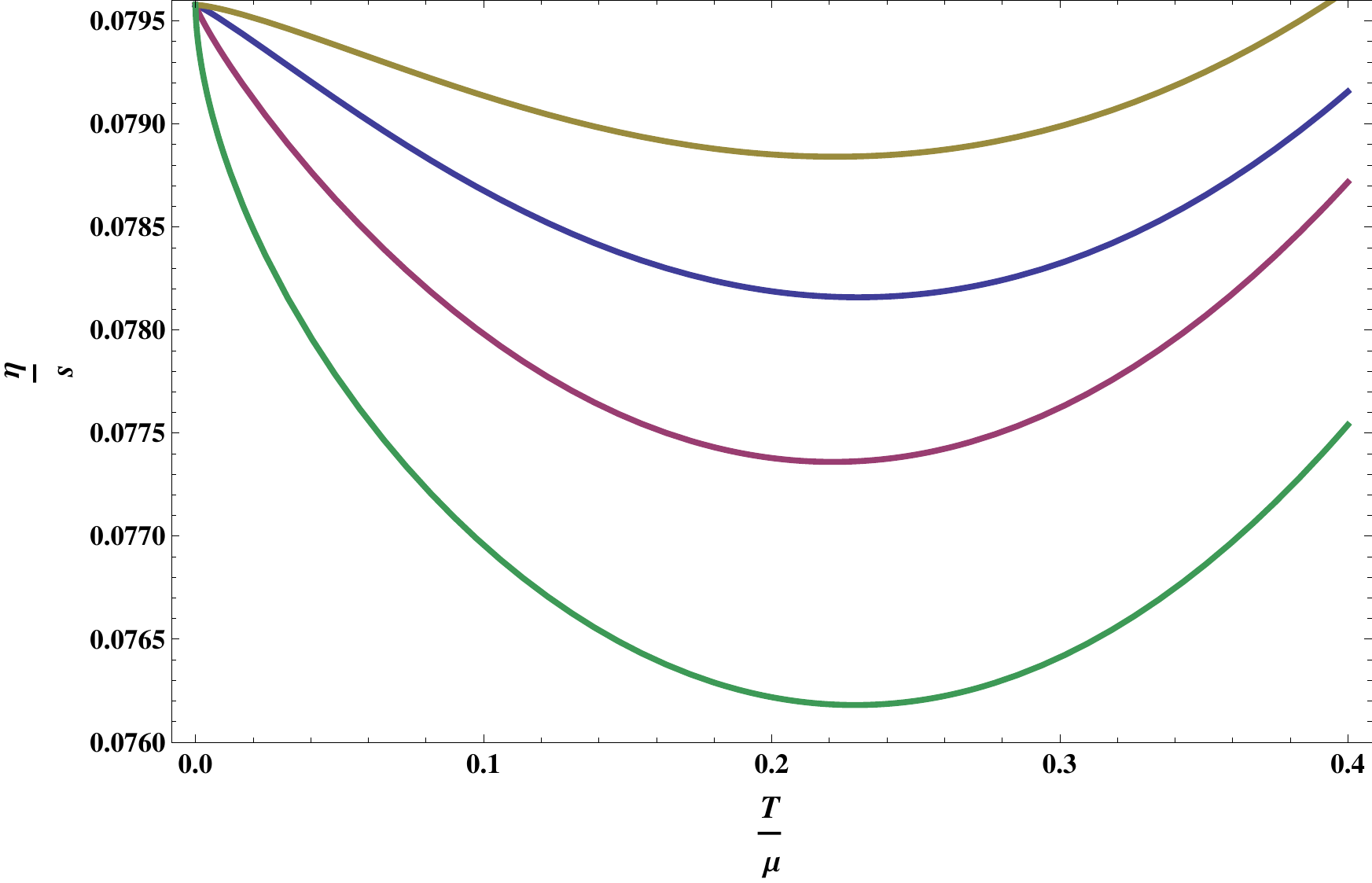}
\caption{$T/\mu$ dependence of $\eta/s$ illustrating existence of minima, reminiscent of data for various liquids in nature.
All curves have the same parameter values of $z=1,$ $\beta_1=0.01,$ and $\beta_2=0.1$
with differing values of $\gamma_1 = \{-2.5,\,-2,\,-1.5,\,-1\}$ and $\gamma_2 = \{-0.2,\,-0.2,\,-0.25,\,-0.75\},$ from top to bottom.}
\label{fig:etas-min}
\end{figure*}

Sample plots of the temperature dependence\footnote{In the plots, we have absorbed the (unknown) constant of proportionality
appearing in (\ref{phi0scaling}) into the values of $\beta_1$ and $\beta_2$, since the latter are free parameters. However,
we have insisted everywhere in the plots that $\beta_1$ and $\beta_2$ are small.}
of $\eta/s$ in the IR can be found in Figures \ref{fig:etas-simple} and \ref{fig:etas-min}.
In Figure \ref{fig:etas-simple} we have chosen to display the simplest dependence of $\eta/s$
on $T/\mu$, in which all parameters except for $z$ and $\beta_2$ are set to zero.
This case corresponds to that described in (\ref{newT}), and the
resulting curves show a simple power law temperature behavior for
$\left(\eta/s -1/4\pi\right)$,
with the power determined by the value of the dynamical exponent $z$.
Thus, one can control how sharply $\eta/s$ varies with temperature by appropriately tuning $z$.

Going back to the more general case described by (\ref{etasLifGen}), in Figure \ref{fig:etas-min} we have fixed instead $z = 3$
as well as the values of $\beta_1$ and $\beta_2$, but have allowed $\gamma_1$ and $\gamma_2$ to vary between each curve.
The parameters have been chosen to allow for minima for $\eta/s$ at some temperature, as observed in many substances in nature.
We emphasize that while the presence of extrema
is a rather typical feature\footnote{Extrema can be easily
obtained by choosing appropriately the signs and exponents of the competing contributions of the $\beta_1$ and $\beta_2$ terms in (\ref{etasLifGen}).}
of the $T/\mu$ dependence of our expression (\ref{etasLifGen}) for $\eta/s$,
the plots we have shown probe only a very small region of parameter space,  and that a much wider range of behavior is possible.
Clearly, for realistic phenomenological applications it would be valuable to construct more concrete models in which at least some of the parameters
governing the temperature dependence can be fixed unambiguously. This is an important question that we plan to address in future work.

\section{Discussion and Summary of Results}
\label{Conclusions}

A question that has emerged quite naturally in the context of the holographic gauge/gravity duality is whether hydrodynamic transport coefficients
depend on the radial direction of the gravitational background used to compute them.
On the field theory side of the duality, the radial dependence is understood as dependence on the RG scale.
It has been known for some time that -- in the AdS/CFT framework -- the shear viscosity $\eta$ can be extracted from quantities which
do \emph{not} depend on the holographic radial coordinate. In other words, it does not run with energy
(see \cite{Eling:2011ct,arXiv:1109.2698} for related recent discussions).

Even though $\es$ doesn't exhibit RG flow, its behavior is not entirely trivial.
It is possible to have constructions in which $\es$ takes on very different values at low and high temperatures (which we refer to as the
IR and UV, respectively),
with a non-trivial temperature flow connecting the two regimes.
One way of achieving such a flow in rotationally symmetric backgrounds is to include higher
derivative corrections in the theory, coupled to a scalar field.
This was done for example in \cite{Buchel:2010wf}, where the temperature dependence of $\es$
-- and the fact that its IR behavior was very different from that in the UV --
was linked to the presence of a non-trivial scalar field controlling the strength of the higher derivative operators,
as well as to the presence of a phase transition.
It is natural to ask, then, whether the running of $\es$ with temperature, and the type of ``decoupling'' of IR from UV physics
observed in \cite{Buchel:2010wf}, can be seen in other setups, and not necessarily involving phase transitions.

With these motivations in mind, in this paper we have chosen to work with a model of
Einstein-Maxwell-dilaton gravity,
whose black brane solutions interpolate smoothly between a near-horizon Lifshitz scaling
in the low-temperature regime and standard AdS scaling at high temperature.
The temperature dependence of the solutions
translates into a temperature flow for $\es$, once higher derivative corrections are turned on.
Thus, this construction provides an explicit setting in which $\es$ exhibits a non-trivial behavior as it varies
as a function of $T/\mu$ from the IR to the UV.
To illustrate some of the features with a simple example, we summarize the case of $\beta_2=\gamma_1=\gamma_2=0$, for which:
\begin{equation}
\etas = \
\begin{cases}
\frac{1}{4\pi} \left[1-8\beta_1 \right] \,,\qquad \qquad {\rm T\gg\mu \; \;\; (UV)} \; ,\\
\frac{1}{4\pi} \left[1-\frac{24}{z+2} \, \beta_1 \right] \,,\qquad \; \, {\rm T\ll\mu \;\; \; (IR) }\; .
\end{cases}
\label{etasuvir}
\end{equation}
When the chemical potential is the dominant scale in the system, the putative field theory
is in the non-relativistic scaling regime,
hence the IR dependence of $\es$ on the dynamical exponent $z$ (or equivalently $\alpha$).
In the relativistic high temperature regime, on the other hand, any knowledge of $z$ is lost -- the shear fluctuation
is now probing a geometry whose near-horizon behavior is AdS-like.
The qualitatively different behavior seen in (\ref{etasuvir}),
depending on the relative hierarchy between the two scales in the theory,
is reminiscent to that of \cite{Buchel:2010wf}.

The simple high-$T$ result $\es = \frac{1}{4\pi} \left[1- 8\beta_1 \right]$ matches that obtained from
curvature-squared corrections in AdS \cite{Kats:2007mq}.
In that case, the couplings of the higher derivative terms could be
related to the central charges of the dual UV CFT via the holographic Weyl anomaly (in our notation $\beta_1 = \frac{1}{8} \frac{c-a}{a}$),
so that one could write $\es$  entirely in terms of UV data.
The situation in our model is quite different: although at high temperatures it is analogous to that of \cite{Kats:2007mq},
in the low-T regime the behavior of $\es$ is no longer determined only by UV quantities,
but it also depends on IR data -- in this case, the dynamical exponent $z$.

In addition to the `simple' $T/\mu$ flow that is visible from (\ref{etasuvir}), describing how $\es$ varies
between its low-$T$ and high-$T$ values, there is another source of temperature dependence, intimately tied to the presence of the
running scalar field.
In fact, in the Liftshitz scaling regime simple dimensional analysis arguments tell us that
\beq
\label{Dim}
e^{-2\alpha\phi_h} \sim \left(\frac{T}{\mu}\right)^{6/z}\, ,
\eq
with $\phi_h$ the horizon value of the scalar. Thus, any factors of $\phi_h$ in the viscosity to entropy ratio will lead
generically to
\emph{additional} sensitivity to temperature.
We can see explicitly the appearance of these new temperature terms by looking at the form of $\es$ in the
$T\ll\mu$ regime, for arbitrary higher derivative couplings:
\begin{equation}\label{etasLifGen1}
\left . \frac{\eta}{s} \, \right\vert_{IR} = \frac{1}{4\pi}\left[1 - \frac{24(1-\frac{3\gamma_1}{\alpha})}{z+2}\; \beta_1 e^{\,\gamma_1\phi_h} +
 \frac{12(z-1)}{\, z+2} \; \beta_2 e^{(\gamma_2-2\alpha) \phi_h} \right]\, .
\end{equation}
The factors of $\phi_h$ will source new temperature dependence through (\ref{Dim}), even when the scalars are not
directly coupled to the higher derivative operators ($\gamma_i=0$).

Interestingly, the terms that generate this additional $T/\mu$ dependence disappear when the higher derivative couplings
are chosen to satisfy the shift-symmetry
\beq
\label{shiftsymm1}
\phi \rightarrow \phi+ \delta\, , \qquad \mathcal A \rightarrow e^{-\alpha \delta} \mathcal A \, ,
\eq
of the two-derivative action.
Specifically, when the dilatonic couplings are such that the symmetry is preserved, \emph{i.e.} $\gamma_1=0$ and $\gamma_2 =2\alpha$,
the expression (\ref{etasLifGen1}) reduces to
\beq
\label{bla}
\left . \frac{\eta}{s} \, \right\vert_{IR} = \frac{1}{4\pi}\left[1 - \frac{24}{z+2}\, \beta_1 +
 \frac{12(z-1)}{z+2} \; \beta_2 \right]\, ,
\eq
with any trace of the additional temperature factors lost.
Moreover, as discussed in detail in Section \ref{nearhorizonexp}, the structure of (\ref{bla}) turns out to be equivalent -- after an
appropriate rescaling of the charge -- to that of a RN-AdS brane.
While this may seem surprising at first, it is easily explained by noting that when the shift symmetry is preserved,
the Lagrangian takes the simpler form
\begin{equation}
{\cal L} = R - 2\Lambda - 2 \left(\nabla \phi\right)^2 - e^{2\alpha\phi} F^2 + L^2\beta_1 \, R_{\mu\nu\rho\sigma}R^{\mu\nu\rho\sigma} +
L^2 \beta_2 \, e^{\, 2\alpha\phi}R_{\mu\nu\rho\sigma}F^{\mu\nu}F^{\rho\sigma}.
\end{equation}
Thus,
the only role of the dilatonic coupling $e^{2\alpha \phi_h}$ at the horizon -- where $\es$ is evaluated --
is to rescale the charge of the $U(1)$ gauge field.

This construction has allowed us to explore some of the features needed to generate a temperature flow
for the transport coefficients of a strongly coupled plasma.
The key ingredients in our model were a non-trivial scalar field profile and different scalings in the IR and in the UV.
The dynamical exponent $z$ and the running of the scalar field
conspired to trigger temperature flow, and to generate a particularly
non-trivial structure for $\es$, with the breaking of the shift symmetry playing a crucial role in giving rise to an additional
source of temperature dependence.
Although it would have been interesting to understand the intermediate temperature regime, here we have restricted our analysis
to the endpoints of the flow, since this was enough to make the points we were after.
Finally, we should note that it would be valuable to extend this work by performing a more systematic -- and more quantitative --
study of the temperature dependence
of the transport coefficients of a strongly coupled plasma,
especially in view of the current applications to the physics of the strongly coupled QGP.
With this context in mind, we emphasize that the key ingredient for achieving non-trivial temperature dependence in the IR
was the existence of a running scalar, and not necessarily the presence of
non-relativistic scaling.
Thus,
similar constructions should conceivably achieve an analogous temperature flow in relativistic settings
more applicable to  QGP-like plasmas.

\section{Acknowledgments}
We are particularly grateful to Ben Burrington for answering patiently all of our questions.
We would like to thank Alex Buchel for valuable input and for comments on the draft.
We also thank Allan Adams, Ibrahima Bah, Umut Gursoy, Sean Hartnoll, Liza Huijse, Jim Liu, Omid Saremi, Julian Sonner and Andy Strominger
for useful discussions at various stages of this work.
S.C. is grateful to KITP for hospitality during the workshop on Holographic Duality and Condensed Matter Physics
while this work was being completed. P.S. would like to acknowledge the hospitality of the George and Cynthia Mitchell Institute for Fundamental Physics and Astronomy at Texas A\&M University where much of this project was initiated. The work of S.C. has been supported by the Cambridge-Mitchell Collaboration in Theoretical
Cosmology, and the Mitchell Family Foundation. The work of P.S. has
been supported by the U.S. Department of Energy under Grant No. DE-FG02-97ER41027.
This research was  supported in part by the National Science Foundation under Grant No. PHY05-51164.

\end{document}